\documentclass[11pt]{article}
\usepackage{amsmath}
\usepackage{amsfonts}
\usepackage{amssymb}
\usepackage{graphicx}
\usepackage{subfigure}
\usepackage{euscript}
\usepackage{color}
\newcommand{\goodgap}{\hspace{\subfigtopskip} \hspace{\subfigbottomskip}}

\def\bea{\begin{eqnarray}}
\def\eea{\end{eqnarray}}

\begin{document}
\begin{center}
\LARGE {\bf Cosmological viability conditions for $f(T)$ dark energy models  }
\end{center}
\begin{center}
 {\bf M. R. Setare\footnote{rezakord@ipm.ir} \\
  N. Mohammadipour\footnote{N.Mohammadipour@uok.ac.ir}}\\
 {\ Department of Science, University of Kurdistan \\
Sanandaj, IRAN.}

 \end{center}
\vskip 3cm
\begin{center}
{\bf{Abstract}}
\end{center}

Recently $f(T)$ modified teleparallel gravity where T is the torsion scalar has been proposed as the natural gravitational alternative for dark
energy. We perform a detailed dynamical analysis of these models and find conditions for the cosmological viability of $f(T)$ dark
energy models as geometrical constraints on the derivatives of these models. We show that in the phase space exists two cosmologically viable trajectory which (i) The universe would start from an unstable radiation point, then pass a saddle standard matter point which is followed by accelerated expansion de sitter point. (ii) The universe starts from a saddle radiation epoch, then falls onto the stable matter era and the system can not evolve to the dark energy dominated epoch. Finally, for a number of $f(T)$ dark energy models were proposed in the more literature, the viability conditions are investigated.

{\bf }

\newpage

\section{Introduction}
Recent cosmological observations indicate that our universe is in accelerated expansion. These observations are those which is obtained by SNe Ia \cite{1}, WMAP \cite{2}, SDSS \cite{3} and X-ray \cite{4}. These observations also suggest that our universe is spatially flat, and consists of
about 70\% dark energy (DE) with negative pressure, 30\% dust matter (cold dark matter plus baryons), and negligible radiation.
In order to explain why the cosmic acceleration happens, many theories have been proposed. The simplest candidate of the dark energy is a tiny positive time-independent cosmological constant $\Lambda$, for which $\omega = -1$. However, it is difficult to understand why the cosmological constant is about 120 orders of magnitude smaller than its natural expectation (the Planck energy density). This is the so-called cosmological constant problem. Another puzzle of the dark energy is the cosmological coincidence problem: why are we living in an epoch in which the dark energy density and the dust matter energy are comparable?. An alternative proposal for dark energy is the dynamical dark energy scenario. The dynamical nature of dark energy, at least in an effective level, can originate from various fields, such is a canonical scalar field (quintessence) \cite{5}, a phantom field, that is a scalar field with a negative sign of the kinetic term \cite{6}, or the combination of quintessence and phantom in a unified model named quintom \cite{7}. In the other hand modified models of gravity provides the natural gravitational alternative for dark energy \cite{8}. Moreover, modified gravity present natural unification of the early time in action and late-time acceleration thanks to different role of gravitational terms relevant at small and at large curvature. Also modified gravity may naturally describe the transition from non-phantom phase to phantom one without necessity to introduce the exotic matter. Among these theories, scalar-tensor theories \cite{9}, $f(R)$ gravity \cite{10} are studied extensively.

Recently a modified gravitational theory, the $f(T)$ gravity, has been proposed \cite{13,14} to explain the acceleration of the cosmic expansion which attracts much attention. In the simple case $f(T)=T$, the $f(T)$ theory can be directly reduced to the Teleparallel Equivalent of General Relativity (TEGR) which was first propounded by Einstein in $1928$ \cite{12}. Similar to the $f(R)$ theories, $f(T)$ theories deviate from Einstein gravity by a function $f(T)$ in the Lagrangian, where T is the so-called torsion scalar. Types of these models have been proposed to explain the late-time acceleration of the cosmic expansion without including the exotic dark energy \cite{13}-\cite{27}.

In this paper we will investigate the cosmologically viable conditions for general $f(T)$ gravity. Wu et. al, \cite{22} and Zhang et. al, \cite{23} in a similar way have investigated the dynamical behaviors of $T+f(T)$ model. They obtained two critical points for radiation and matter dominated, respectively and a critical line for dark energy dominated. Then for a concrete power-law model, they obtained that dynamical system has a stable de sitter phase along with an unstable radiation dominated phase and an unstable matter dominated one.

However, in our study for general $f(T)$ model we obtain three critical continuous lines corresponding to de sitter points $(\Omega_{m}=0, \Omega_{r}=0, \Omega_{DE}=1, \omega_{eff}=-1)$, scaling solution with matter $(\Omega_{m}=1, \Omega_{r}=0, \Omega_{DE}=0, \omega_{eff}=0)$ and scaling solution with radiation $(\Omega_{m}=0, \Omega_{r}=1, \Omega_{DE}=0, \omega_{eff}=\frac{1}{3})$, respectively. we find the general conditions for the cosmological viablility of these models as geometrical constraints on the derivatives their and obtain two kinds of cosmological viable trajectory for general $f(T)$ model: $(i)$ we have a saddle matter phase which is followed by a de sitter epoch along with an unstable radiation era $(ii)$ the system will stay in the stable matter dominated phase and can not evolve to dark energy dominated which would start from a saddle radiation era.

The paper is organized as follows, in the following section, we review $f(T)$ theories and in order to study the dynamics of these models, we have to rewrite the field equations into a $4$ dimensional autonomous system which once is a constant. using the method \cite{11}, we construct the curve $m(r)$ which $m=\frac{Tf''(T)}{f'(T)}$ and $r=-\frac{T f'(T)}{f(T)}$, where a prime denotes derivative with respect to the torsion scalar $T$. In section $\ref{3}$, we drive the critical lines with their stabilities then in the $(r, m)$ plane, we obtain conditions for the cosmological viable of $f(T)$ dark energy models. Section \ref{4} is based on the numerical analysis for a number of $f(T)$ models to validity the analytical results presented in the section $\ref{3}$. Finally, in Section \ref{5}, we present our conclusions.

\section{$f(T)$ dark energy models  }

In this section firstly, we briefly review the Teleparallel gravity so-called $f(T)$ gravity in the spatially flat FRW universe. Then we obtain the autonomous equations to study cosmological dynamics of $f(T)$ gravity models.
\begin{center}
{\bf{A.Field equations and definitions}}
\end{center}
We start with the generic form of the action of Teleparallel gravity as
\begin{equation}\label{1}
 S=\frac{1}{k^{2}}\int edx^4(f(T)+L_{r}+L_{m}),
\end{equation}
where $k^{2}=8\pi{\cal G}$, $e=\sqrt{-g}=det(e^{i} _{\mu})$ and $T$ is the torsion scalar, also $L_{r}$ and $L_{m}$ are the Lagrangian density of the radiation and the matter, respectively.

In the $f(T)$ framework, in the spatially flat FRW universe $e^{i} _{\mu}$, the veribein field is related to the metric as
\begin{eqnarray}\label{2}
g_{\mu\nu}=\eta_{ij} e^{i}_{\mu} e^{i}_{\nu},
\end{eqnarray}
here $\mu, \nu$ are the coordinate indices on the manifold while $i, j$ are the coordinate indices for the tangent space of the manifold which all indices run over $0, 1, 2, 3$, also $\eta_{ij}$=diag(1, -1, -1, -1).

Varying the action Eq.(\ref{1}) with respect to the veribein field leads to the equations \cite{14}
\begin{eqnarray}\label{3}
  S_i^{~\mu\nu}\partial_{\mu}
(T)f_{TT}(T)+\Big[e^{-1}\partial_{\mu}(eS_i^{~\mu\nu})-e_i^{\lambda}T^{\rho}_{~\mu\lambda}S_{\rho}^{~\nu\mu}\Big]f_T(T)\\
\nonumber
+\frac{1}{4}e_i^{\nu}f(T)=\frac{k^2}{2}e_i^{~\rho}T_{\rho}^{~\nu},~~~~~~~~~~~~~~~~~~~~~~~~~~~~~~~~~~~~~~~~~~
\end{eqnarray}
Here a prime denotes differentiations with respect to the torsion scalar $T$. Also $S_{i}^{\mu\nu}=e_{i}^{\rho}S_{\rho}^{\mu\nu}$ and $T_{\mu\nu}$ is the matter energy- momentum tensor.\\
In the spatially flat FRW metric $g_{\mu \nu}$=diag(-1, a(t), a(t), a(t)), the set of field equations Eq.(\ref{3}) for $i=0=\nu$ reduce to \cite{14}
\begin{equation}\label{4}
12 H^{2}f'(T)+f(T)=2k^{2}(\rho_{r}+\rho_{m}).
\end{equation}
and for $i=1=\nu$ lead to
\begin{equation}\label{5}
48H^{2}\dot{H}f''(T)-(12H^{2}+4\dot{H})f'(T)-f(T)=2k^{2}(p_{r}+p_{m}).
\end{equation}
and torsion scalar as a function of the Hubble parameter $H=\frac{\dot{a}}{a}$
\begin{equation}\label{6}
T=-6H^{2}.
\end{equation}
Where a dot represents a derivative with respect to the cosmic time $t$ also, $(\rho_{r}, p_{r})$ and $(\rho_{m}, p_{m})$ are the ($total$ $energy$ $density$, $pressure$) of the radiation and the matter inside the universe, respectively.

Using Eqs.(\ref{4}, \ref{5}), one can obtain the modified Friedmann equations as follows
\begin{eqnarray}\label{7}
3H^{2}&=&k^{2}(\rho_{r}+\rho_{m}+\rho_{T}),\\
2\dot{H}+3H^{2}&=&-k^{2}(p_{r}+p_{m}+p_{T}),
\end{eqnarray}
with
\begin{eqnarray}\label{9}
\rho_{T}&=&\frac{1}{2k^2}(2Tf'(T)-f(T)-T),\\
p_{T}&=&-\frac{1}{2k^2}[4\dot{H}(-2Tf''(T)-f'(T)+1)]-\rho_{T}.
\end{eqnarray}
Here $\rho_{T}$ and $p_{T}$ are the torsion contributions to the total energy density and pressure, respectively.\\
The continuity equations of these energy densities $\rho_{r}$, $\rho_{m}$, $\rho_{T}$ are:
\begin{eqnarray}\label{11}
0&=&\dot{\rho_{r}}+4H\rho_{r},\\
0&=&\dot{\rho_{m}}+3H\rho_{m},\\
0&=&\dot{\rho_{T}}+3H(\rho_{T}+p_{T}).
\end{eqnarray}

From Eqs.(\ref{9}), (10), we can define gravitationally induced form of dark energy density $\rho_{T}=\rho_{DE}$ and pressure $p_{T}=p_{DE}$.
The equation of state parameter is defined as
\begin{eqnarray}\label{16}\
\omega_{DE}=\frac{p_{DE}}{\rho_{DE}}=-\frac{-8\dot{H}T f''(T)+(2T-4\dot{H})f'(T)-f(T)+4\dot{H}-T}{2Tf'(T)-f(T)-T},
\end{eqnarray}
and for a $f(T)$ (DE) dominated universe, one can obtain the effective equation of state
\begin{eqnarray}\label{17}
\omega_{eff}=-1-\frac{2\dot{H}}{3H^2}\frac{2Tf''(T)+f'(T)-1}{2f'(T)+\frac{f(T)}{6H^2}-1}.
\end{eqnarray}
\begin{center}
{\bf{B.Autonomous equations for $f(T)$ dark energy}}
\end{center}

To study the dynamics of a general $F(T)$ model as a dynamical system, we introduce the dimensionless variables as follows
\begin{eqnarray}\label{18}
x_{1}&=&\frac{k^2\rho_{r}}{3H^2},\\
x_{2}&=&-2f'(T),\\
x_{3}&=&-\frac{f(T)}{6H^2},\\
x_{4}&=&-\frac{T}{6H^2}=1.
\end{eqnarray}
Using the above relations we can rewrite Eq.(\ref{7}) as the following equation
\begin{eqnarray}\label{22}
\Omega_{m}=1-x_{1}-x_{2}-x_{3}-x_{4},
\end{eqnarray}
with the density parameters
\begin{eqnarray}\label{23}
\Omega_{i}=\frac{k^2\rho_{i}}{3H^2} \hspace {2cm} i=radiation, matter\hspace{2mm} and\hspace{2mm} DE.
\end{eqnarray}
One can rewrite Eqs.(7-11) as the following equations of motion
\begin{eqnarray}\label{24}
\frac{dx_{1}}{dN}&=&-2x_{1}(2+\frac{\dot{H}}{H^2}),\\
\frac{dx_{2}}{dN}&=&2mx_{2}\frac{\dot{H}}{H^2},\\
\frac{dx_{3}}{dN}&=&-(x_{2}+2x_{3})\frac{\dot{H}}{H^2},
\end{eqnarray}
with
\begin{eqnarray}\label{27}
\frac{\dot{H}}{H^2}=\frac{-3x_{2}-3x_{3}+x_{1}}{(2m+1)x_{2}}.
\end{eqnarray}

The autonomous dynamical system Eqs.(22-24) is the general dynamical system that describes the cosmological dynamics of $f(T)$ models.\\
Where $N=\ln(\frac{a}{a_{i}})$ and\footnote{$a_{i}$ is the initial value of the scale factor.}
\begin{eqnarray}\label{28}
m=\frac{Tf''(T)}{f'(T)},
\end{eqnarray}
\begin{eqnarray}\label{29}
r=-\frac{T f'(T)}{f(T)}=\frac{x_{2}}{2x_{3}}.
\end{eqnarray}
From Eq.(\ref{29}) one can obtain torsion scalar $T$ as a function of $r=\frac{x_{2}}{2x_{3}}$ then, we can say that $m$ is a function of $r$ and $m=m(r)$.\\
The effective equation of state Eq.(\ref{17}) can be rewritten in terms of $x_{i}$ which are defined in Eqs.(16-19) as
\begin{eqnarray}\label{30}
\omega_{eff}=-1-\frac{1}{3}(\frac{-3x_{2}-3x_{3}+x_{1}}{(2m+1)x_{2}})(\frac{(2m+1)x_{2}+2}{x_{2}+x_{3}+1}).
\end{eqnarray}
Eqs.(22-24) show that the results of our analysis depend on $x_{i}$ Eqs.(16-19) and $m(r)$ Eq.(\ref{28}). Therefore, given a form of $f(T)$, we obtain a function $m(r)$. For instance, the power-law model $f(T)=T+T^{n}$ corresponds to $m(r)=n\frac{r+1}{r}$ in the $(r, m)$ plane, similar to $f(R)$ and $f(G)$ \cite{11}. Also given a form of $m$ as a function of $r$ that is cosmologically viable, one may use Eqs.(\ref{28}), (\ref{29}) reconstructing the $f(T)$ viable model.

On the other hand, from Eqs.(\ref{28}), (\ref{29}) we can derive the following relation for $r$:
\begin{eqnarray}\label{31}
\frac{d}{dN}r=\frac{\dot{T}}{HT}r(m+r+1).
\end{eqnarray}
It shows that $\frac{d}{dN}r=0$ if $r=0$ and $\frac{\dot{T}}{HT}(m+r+1)$ does not diverge or $m+r+1=0$ and $\frac{\dot{T}}{HT}r$ does not diverge. This means that the evolution of the system along the curve $m(r)$ stops for any intersection points between $r=0$ (the $m$ axis) or critical line $m=-r-1=0$ and the curve $m(r)$.

\section{Phase-space analysis for general $f(T)$ models  }

In order to study the dynamical behavior of the system Eqs.(22-24), by sitting $\frac{dx_{i}}{dN}=0$ we find three continuous lines of critical points with their properties in the following relations as:
\begin{eqnarray}\label{32}
L_{1} : (x_{1}=0,\hspace{1cm} x_{2}=-x_{3},\hspace{1cm} x_{3}=x_{3}),~~~~~~~~~~~~~~~~~~~~~~\\
\nonumber
\Omega_{m}=0,\hspace{1cm}  \Omega_{r}=0,\hspace{1cm}  \Omega_{DE}=1,\hspace{1cm}  \omega_{eff}=-1,~~~~~~~~
\end{eqnarray}

\begin{eqnarray}\label{33}
L_{2} : (x_{1}=0,\hspace{1cm} x_{2}=-2x_{3},\hspace{1cm} x_{3}=x_{3},\hspace{1cm}m=0),~~~\\
\nonumber
\Omega_{m}=x_{3},\hspace{1cm}  \Omega_{r}=0,\hspace{1cm}  \Omega_{DE}=1-x_{3},\hspace{1cm}  \omega_{eff}=0,~~
\end{eqnarray}

\begin{eqnarray}\label{34}
L_{3} : (x_{1}=x_{3},\hspace{1cm} x_{2}=-2x_{3},\hspace{1cm} x_{3}=x_{3},\hspace{1cm}m=0),\\
\nonumber
\Omega_{m}=0,\hspace{1cm}  \Omega_{r}=x_{3},\hspace{1cm}  \Omega_{DE}=1-x_{3},\hspace{1cm}  \omega_{eff}=\frac{1}{3}.
\end{eqnarray}

It is worth noting that for $L_{1}$, we have $r=-\frac{1}{2}$ and for $L_{2}$ and $L_{3}$ , we have $r=-1$ corresponding to the definition of Eq.(\ref{29}). Also for $L_{2}$ and $L_{3}$ , we have $m=0$ so, for these critical lines:
\begin{eqnarray}\label{35}
m(r=-1)=0.
\end{eqnarray}

Thus for critical lines $L_{2}$ and $L_{3}$ we have a critical point ($r=-1$, $m=0$) in the $(r, m)$ plane which the system stops in
 this point in agreement with the critical line $m=-r-1$. Note that $L_{2}$ and $L_{3}$ corresponds to scaling solutions of the dark energy
 with matter and radiation respectively. Hence in matter or radiation dominated era the $m(r)$ curve should pass through the critical point
  ($r=-1$, $m=0$) in the $(r, m)$ plane. Also $L_{1}$ corresponds to de-Sitter points $(\dot{H}=0)$ which the $m(r)$ curve goes through the critical point
   ($r=-\frac{1}{2}$, $m=-\frac{1}{2}$).\\

   The dynamical behavior of $f(T)$ dark energy models have been investigated in \cite{22}, \cite{23}. Wu et. al, \cite{22} have obtained a critical line for
   the effective dark energy dominated era and two points for matter and radiation dominated. We will see later that the two points obtained for matter and radiation dominated
    are in agreement with the matter dominated point and the radiation dominated point in critical lines $L_{2}$ and $L_{3}$, respectively.\\
Now we shall consider the properties of any point along the critical lines in turn and we study the stability conditions of these points.
\begin{itemize}
\item $P_{1}$ : \emph{de-Sitter point}\\
$(x_{01}=0,\hspace{1mm} x_{02}=-x_{03},\hspace{1mm} x_{03}=x_{03})$

Point $P_{1}$ is characterized by $\omega_{eff}=-1$, $\Omega_{DE}=1$ that corresponds to de-Sitter solutions $(\dot{H}=0)$ for any point along the critical line $L_{1}$ and for any value of $x_{03}$. The matrix perturbation eigenvalues for these points are
\begin{eqnarray}\label{36}
-4,\hspace{5mm}-\frac{3}{2}\pm\frac{2}{(2m+1)}\sqrt{\frac{m(m-1)}{(2m+1)}+\frac{8m^{3}+9m^{2}+3m+\frac{1}{4}}{(2m+1)^2x_{03}}},
\end{eqnarray}
with $m_{ds}=m(r=-\frac{1}{2})=-\frac{1}{2}$. It is worth noting that the de-Sitter point along $L_{1}$ is
 stable and an attractor solution when $m_{ds}\rightarrow(-\frac{1}{2})^{-}$, because the last two eigenvalues
 are complex and have negative real parts.

 Note that, for any value of $x_{03}\neq0$ we have an attractor de sitter point with eigenvalues 0, -3, -4 in agreement with \cite{22}.
\item $P_{2}$ : \emph{scaling solutions with matter point}\\
\hspace{6cm}$(x_{01}=0,\hspace{1mm} x_{02}=-2x_{03},\hspace{1mm} x_{03}=x_{03})$

Point $P_{2}$ corresponds to scaling solutions with the density parameters ratio $(\frac{\Omega_{m}}{\Omega_{DE}}=\frac{x_{03}}{1-x_{03}})$.
The eigenvalues for these points are given by
\begin{eqnarray}\label{37}
0,\hspace{1cm}-1,\hspace{1cm}3(1+m_{,r}),
\end{eqnarray}
here $m_{,r}=\frac{dm}{dr}$. The stability of $P_{2}$ depends on $m_{,r}$.
Considering $x_{03}=1$ this solution leads to $P_{2m}=(0,\hspace{1mm} -2,\hspace{1mm} 1)$ that represents a standard matter epoch with $\Omega_{m}=1$, $\omega_{eff}=0$ and the above eigenvalues. Hence the necessary condition for $P_{2}$ to have a standard matter epoch is $m(r=-1)=0$ and $P_{2}$ is a saddle point when
 \begin{eqnarray}\label{38}
m_{,r}(r=-1)>-1.
\end{eqnarray}
The above relation is the necessary condition to have a standard matter epoch which evolve to dark energy dominated era.
However, this point is stable provided that $m_{,r}(r=-1)\leq-1$ .
This means that the evolution of the system from this point stops and can not move away from it and the system does not evolve to dark energy dominated phase.

\item $P_{3}$ : \emph{scaling solutions with radiation point}\\
\hspace{6cm}$(x_{01}=x_{03},\hspace{1mm} x_{02}=-2x_{03},\hspace{1mm} x_{03}=x_{03})$

At this point, dark energy mimics the evolution of radiation which gives the constant ratio $(\frac{\Omega_{r}}{\Omega_{DE}}=\frac{x_{03}}{1-x_{03}})$ and the point $P_{3r}=(1,\hspace{1mm} -2,\hspace{1mm} 1)$ represents the dominated radiation era with $\Omega_{r}=1$ and $\omega_{eff}=\frac{1}{3}$.
\begin{eqnarray}\label{39}
0,\hspace{1cm}1,\hspace{1cm}4(1+m_{,r}).
\end{eqnarray}
Note that to have an unstable radiation point $m_{,r}(r=-1)>-1$.

It is worth noting that, $P_{3r}$ and $P_{2m}$ points are in agrement with radiation dominated point $A$ and
 matter dominated point $B$ in \cite{22}, respectively. They have obtained only two points as radiation and matter dominated but, we have obtained two critical lines
 as scaling solutions with radiation point and scaling solutions with matter point.

We know that a cosmologically viable trajectory would start from the radiation era, then pass from a matter dominated epoch followed by an accelerated expansion which in the phase space would start from $P_{3}$ then pass the matter dominated point $P_{2}$ and at last land at the stable point $P_{1}$. In the $(r, m)$ plane, in order to have $P_{3}$ and $P_{2}$ the curve $m(r)$ should pass the point $(r=-1, m=0)$ and the criterion for the existence of de-Sitter point $P_{1}$ is given by passing the curve $m(r)$ from point $(r=-\frac{1}{2}, m=-\frac{1}{2})$.
\end{itemize}

\section{Specific cases and numerical results }

In this section we study the cosmological viable trajectory for several $f(T)$ models by writing $m$ as a function of $r$ and we obtain most of the relevant properties of these models.
\begin{center}
{\bf{I. $f(T)=T-\frac{\alpha}{(-T)^{n}}$}}
\end{center}
\begin{figure}
\centering
\includegraphics[width=8cm]{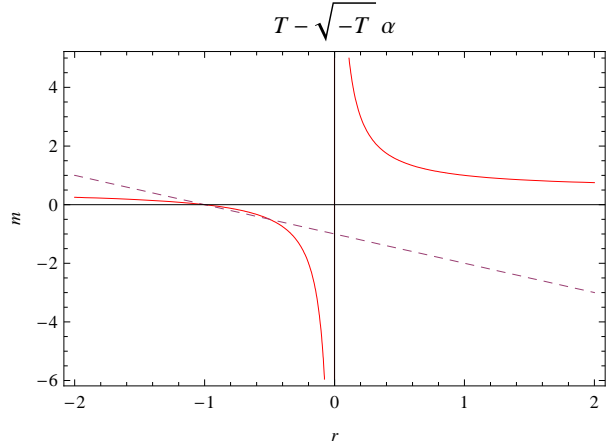}\\
\caption{ The $m(r)$ curve (red line) for power-law model which satisfies the conditions of cosmological viable trajectory. the curve goes through to the radiation and matter point $(r=-1, m=0)$ and the attractor de sitter point $(r=-\frac{1}{2}, m=-\frac{1}{2})$ and $m_{,r}(r=-1)=-\frac{1}{2}>-1$.   }\label{fig: 2}
\end{figure}

This model was proposed in \cite{21}-\cite{24} to explain the late-time accelerated expansion without including the dark energy component.
Also Bengochea et. al, \cite{14} have investigated the observational information for this model by using the most recent SN Ia+BAO+CMB data they found that the values lie in the ranges $n\in[0.23, 0.03]$ and $\Omega_{m}\in[0.25, 0.29]$. The model presents radiation era, matter era
 and late acceleration phases as the last three phases of
 cosmological evolution in standard model. \\
Using Eqs.(\ref{28}), (\ref{29}) one can obtain $m(r)=-n(1+\frac{1}{r})$ which is independent of $\alpha$.
Note that the necessary condition for the existence of the radiation and matter points $m(r=-1)=0$ satisfy by this model. $m(r=-\frac{1}{2})=-\frac{1}{2}$ is the necessary condition to have a stable de sitter point then $n=-\frac{1}{2}$ gives a stable de sitter point corresponding to $f(T)=T-\alpha \sqrt{-T}$ which was proposed in \cite{25,26} explaining the accelerated expansion at late-time. The derivative $m$ with respect to the $r$ is given by
\begin{eqnarray}\label{40}
m_{,r}=n\frac{1}{r^2}.
\end{eqnarray}
Hence for this model $m_{,r}(r=-1)=-\frac{1}{2}>-1$ satisfying the condition Eq.(\ref{38}) to have the matter epoch which is followed by the de sitter dark energy dominated and is shown in Fig.\ref{fig: 2}.
\begin{table}[t]
\begin{center}
\begin{tabular}{|c|c|c|c|c|c|}
\hline
$f(T)$  & $m(r)$ & $m_{,r}$ & $m(r)=0$ & $m_{,r}>-1$ & $m(r)=-\frac{1}{2}$\\
~ & ~ & ~ & ~ & ~ & ~ \\
models & ~ & ~ & $r=-1$ & $r=-1$ & $r=-\frac{1}{2}$ \\
\hline \hline
~ & ~ & ~ & ~ & ~ & ~ \\
$T^{a}e^{bT}$ & $\frac{a-r^{2}}{r}$ & $-\frac{a+r^{2}}{r^{2}}$ &$a=1$  &$a>-2$  & $a=\frac{1}{8}$ \\
~ & ~ & ~ & ~ & ~ & ~ \\
$T^{a}e^{\frac{b}{T}}$ & $-\frac{r^2+2r+a}{r}$ & $\frac{a}{r^2}-1$ &$a=1$  &$a>0$  & $a=\frac{1}{8}$\\
~ & ~ & ~ & ~ & ~ & ~ \\
$T^{a}(\ln(\gamma T))^{b}$ & $\frac{1-b}{b}r^2+\frac{a^2}{br}$ & $-1+\frac{r^2-1}{br}$ &$a=\frac{11\pm 6\sqrt{3}}{13}$&$-1$ &$a=\frac{11\pm 6\sqrt{3}}{13}$  \\
~ & ~ & ~ & ~ & ~ & ~ \\
 &$+\frac{2a-b}{b}$  &  &  &  &  \\
~ & ~ & ~ & ~ & ~ & ~ \\
\hline
\end{tabular}
\end{center}
\caption{Cosmological viability conditions on a number of $f(T)$ models. Columns are as follows\,: 1. $f(T)$ models; 2., 3. Curve $m(r)$
 and its first derivative $m_{,r}$ corresponds to these models; 4. The necessary condition to have radiation and standard matter era; 5.
  The necessary condition to have a standard matter epoch which is followed by the dark energy dominated phase, 6. The necessary condition to
   have a stable de sitter point.}
\label{tab: 1}
\end{table}
\begin{center}
{\bf{II. Exponential model and logarithmic model}}
\end{center}
\begin{figure*}
\centering
\subfigure[\label{fig1da} ]{\includegraphics[width=5cm]{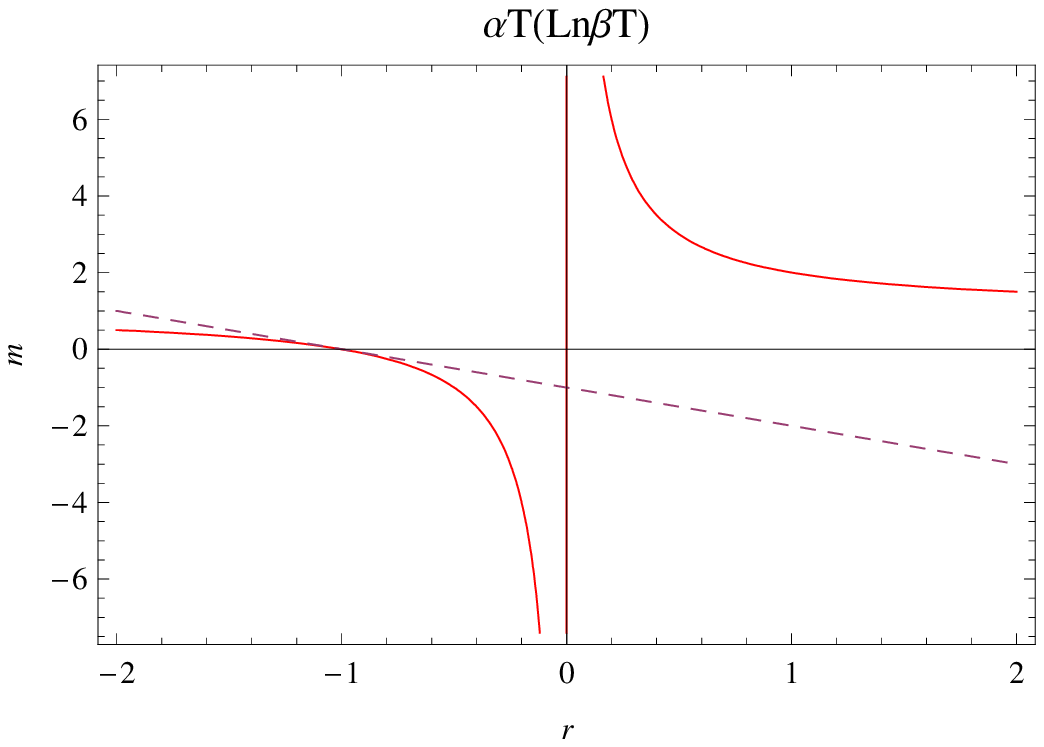}} \goodgap
\subfigure[\label{fig1db} ]{\includegraphics[width=5cm]{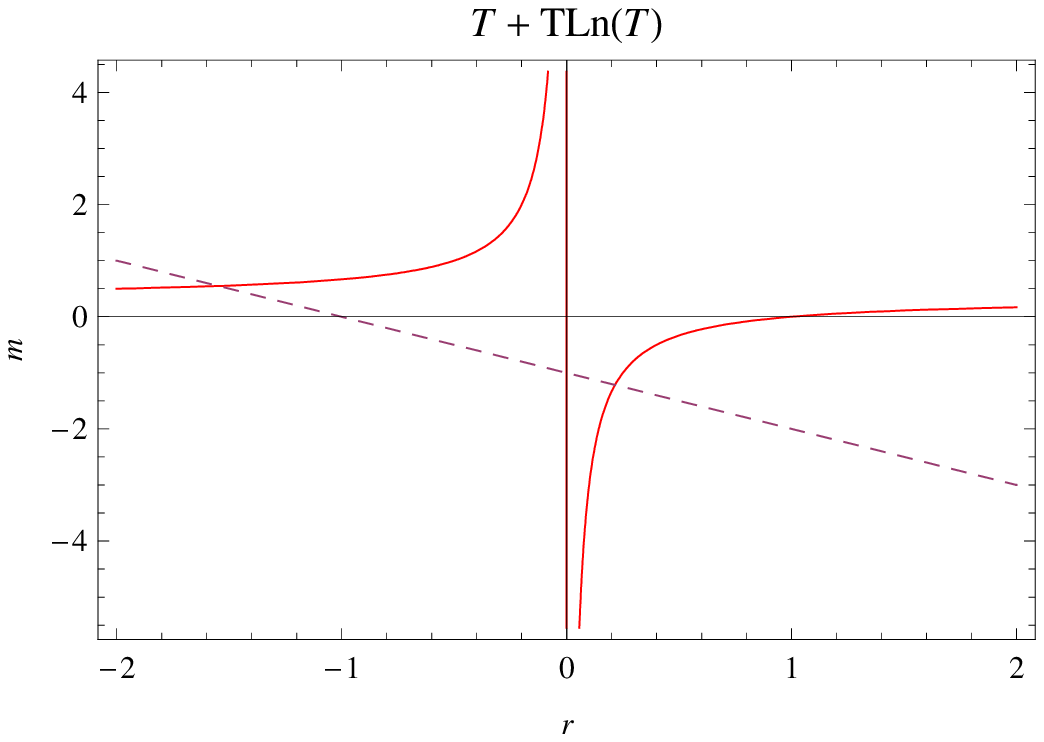}} \goodgap
\subfigure[\label{fig1dc} ]{\includegraphics[width=5cm]{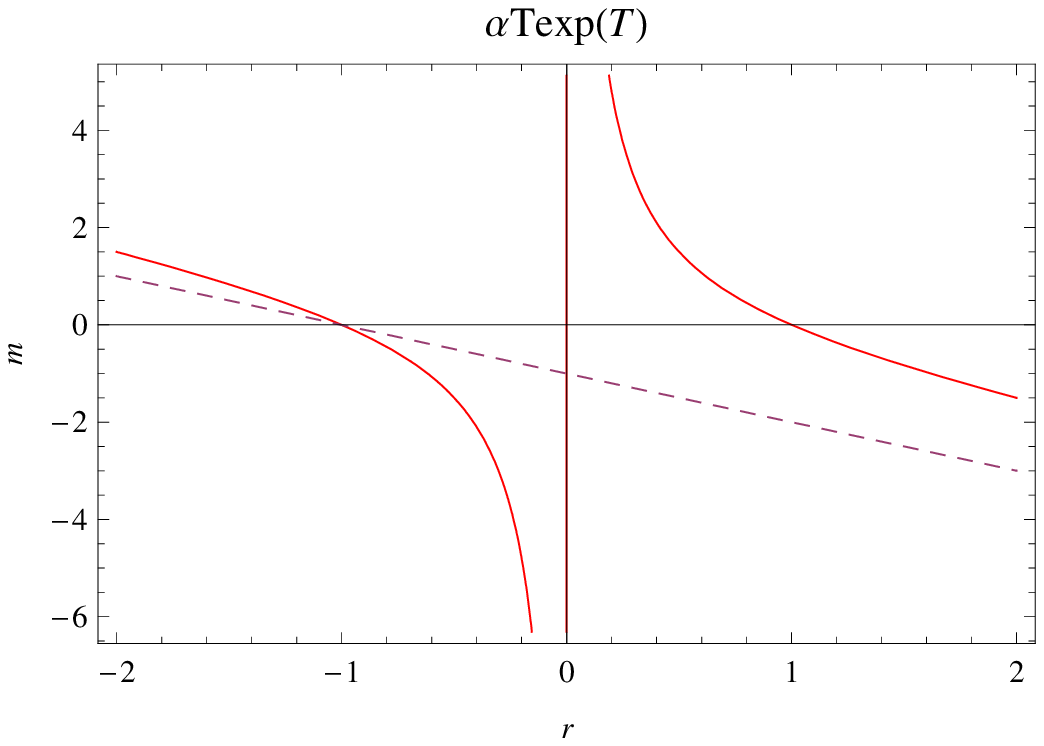}} \goodgap
\subfigure[\label{fig1dd} ]{\includegraphics[width=5cm]{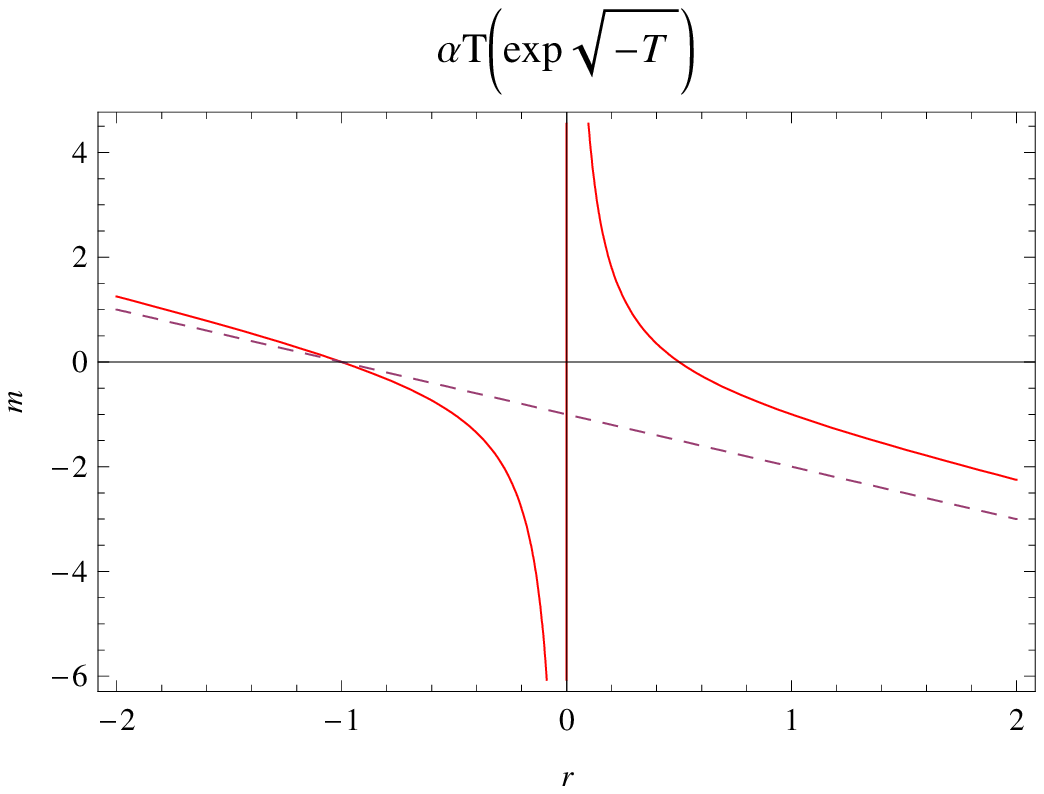}} \goodgap
\subfigure[\label{fig1de} ]{\includegraphics[width=5cm]{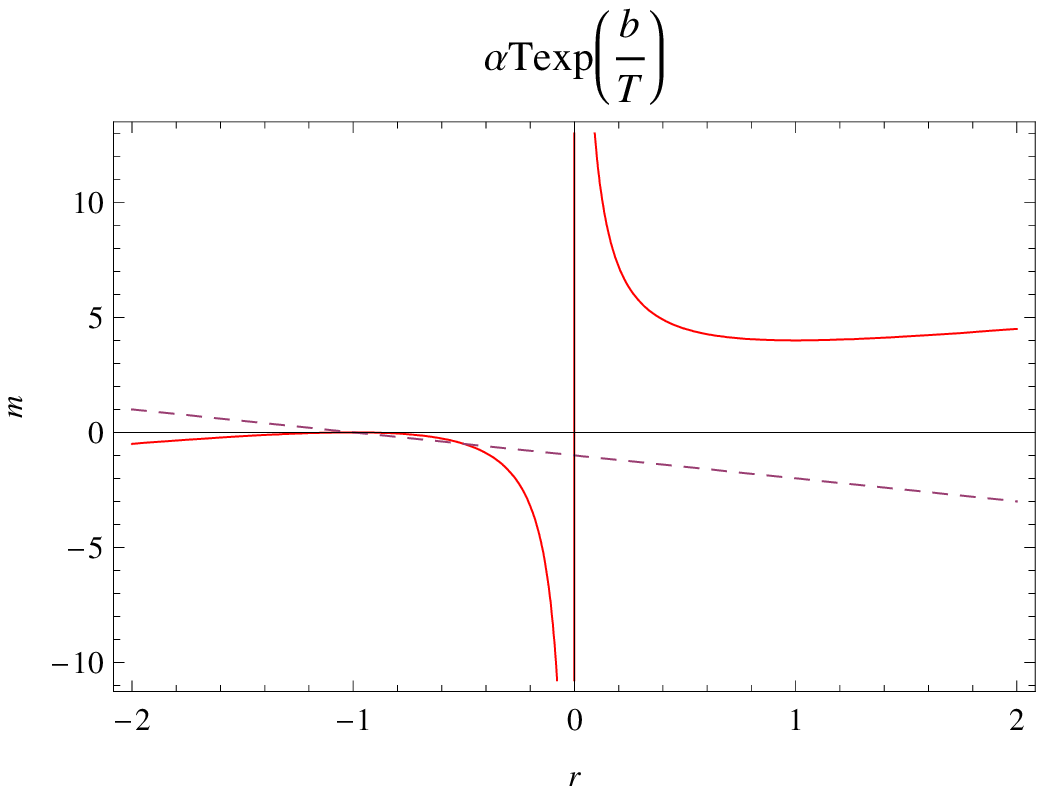}} \goodgap
\subfigure[\label{fig1df} ]{\includegraphics[width=5cm]{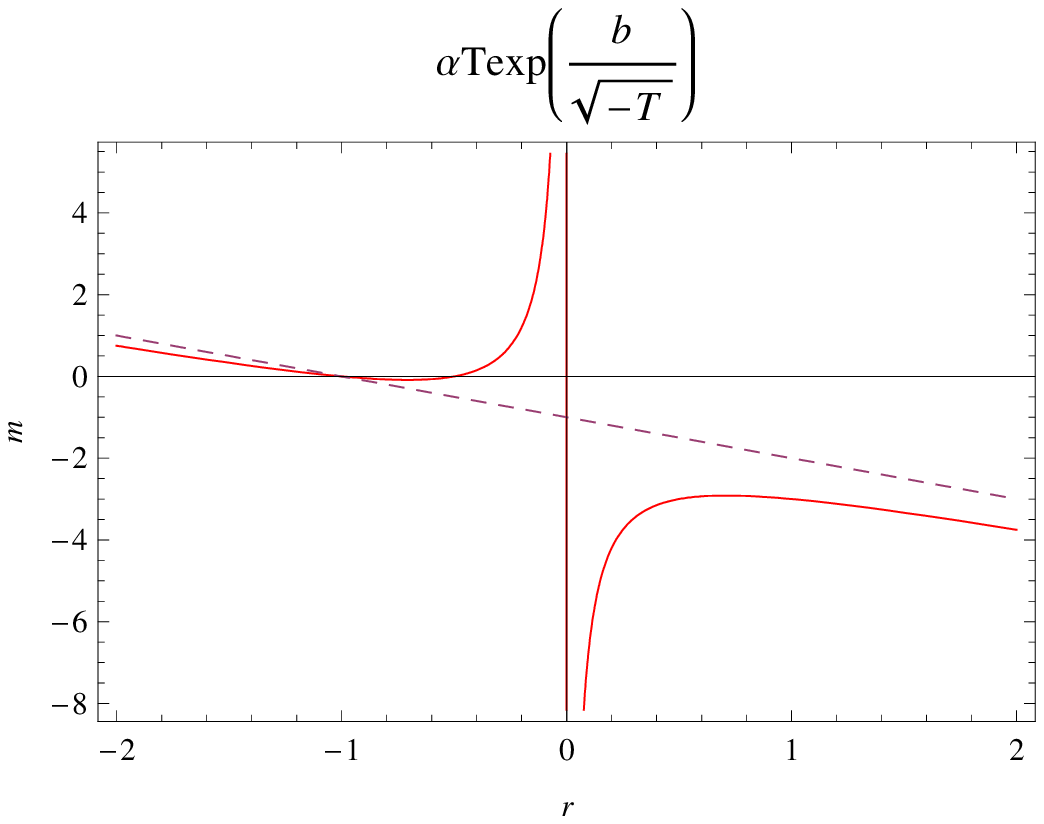}} \goodgap \\
\caption{This figure shows several possible $m(r)$ curves (red lines) for a number of $f(T)$ models. Only the case $b$: $f(T)=T+T\ln T$ the $m(r)$ curve does not stop the critical line $m(r)=-r-1$ at $r=-1$ and these models do not have the standard matter and the radiation era. In the other cases, the $m(r)$ curves intersect the critical line at the matter point $(r=-1, m=0)$ but in all cases the condition Eq.(\ref{38}) does not satisfy to have the standard matter era which is followed by the de sitter dark energy dominated and system stay in the stable matter point where the values of $m_{,r}$ at $r=-1$ are ($-1, -2$ and $-\frac{3}{2}$) for the cases of $a, c$ and $d$, respectively which are smaller than $-1$. For the cases $e$ and $f$ the values of $m_{,r}$ at $r=-1$ are larger than $-1$ but do not pass through the de sitter point.}
\label{fig: 1}
\end{figure*}

In this section, we consider three cases of the exponential model and the logarithmic model which are similar with those of $f(R)$ models
and we can compare them with the models proposed in $f(R)$ theories \cite{11}. These models have been presented in \cite{23}-\cite{26} and
we examine the conditions for the cosmological viability of them. The numerical results are summarized in Table \ref{tab: 1}.

We see that, two exponential models have a matter epoch which is followed by the de sitter point similar to $f(R)=Rexp(\frac{1}{R})$ in $f(R)$
 theories \cite{11} (see Fig.\ref{fig: 1}).The phenomenological model with a logarithmic form $(m_{,r}=-1)$ can not satisfy the condition Eq.(\ref{38}). In this case,
 unlike the standard model, system stays in standard matter era and does not land to the stable de sitter point. These models are ruled out by the
 history of our universe(see Fig.\ref{fig: 1}).

\section{ Conclusion }

The $f(T)$ theory, obtained from generalizing teleparallel gravity, is an alternative modified gravity to explain
 the present cosmic acceleration without including the dark energy. In the present work we investigated the dynamical behavior
 of the recently proposed scenario of $f(T)$ dark energy. Performing a detailed phase-space analysis of $f(T)$ dark energy models,
 we obtained three critical lines corresponding to the de sitter point $(\Omega_{m}=0, \Omega_{r}=0, \Omega_{DE}=1, \omega_{eff}=-1)$,
 scaling solution with matter $(\Omega_{m}=1, \Omega_{r}=0, \Omega_{DE}=0, \omega_{eff}=0)$ and scaling solution with radiation
 $(\Omega_{m}=0, \Omega_{r}=1, \Omega_{DE}=0, \omega_{eff}=\frac{1}{3})$. We found conditions for the cosmological viability of
 these models as geometrical constraints on the derivatives of these models $(r=-\frac{T f'(T)}{f(T)}, m=\frac{Tf''(T)}{f'(T)})$.
 Having de sitter point led to passing the curve $m(r)$ into point $(r=-\frac{1}{2}, m=-\frac{1}{2})$ and to have radiation and
 matter point the curve $m(r)$ should pass through the point $(r=-1, m=0)$. The stabilities for these points were studied and we obtained
 two kinds of cosmologically viable trajectory which $(i)$ the universe would start from an unstable radiation point, then pass a saddle
 standard matter point which would be followed by accelerated expansion de sitter point. $(ii)$ The universe started from a saddle radiation epoch,
 then fell onto a stable matter era and the system could not evolve to the dark energy dominated epoch. Finally, we investigated the dynamical
 behaviors for a number toy models of $f(T)$ dark energy which were proposed in the more literature, recently. These models can exhibit the radiation era, the
  matter era and the late acceleration also, we can compare them with those of $f(R)$ theories.

\end{document}